\documentclass[11pt,letterpaper]{article}
\usepackage[T1]{fontenc}
\usepackage[utf8]{inputenc}
\usepackage{lmodern}
\usepackage[margin=1in]{geometry}
\usepackage{amsmath,amssymb,graphicx,booktabs,longtable,array}
\usepackage{microtype}
\usepackage[round,authoryear]{natbib}

\usepackage{caption}
\usepackage[hidelinks]{hyperref}
\usepackage{url}
\usepackage{placeins}
\newcommand{\vct}[1]{\boldsymbol{#1}}
\newcommand{\mat}[1]{\mathsf{#1}}
\newcommand{\dd}{\,\mathrm d}
\newcommand{\qunit}{\ensuremath{\mu\mathrm W\,\mathrm m^{-2}}}
\newcommand{\punit}{\ensuremath{\mathrm{eV}\,\mathrm{cm}^{-3}}}

\graphicspath{{figures/}}
\title{Constraining electron heat flux from reduced measurements}
\hypersetup{pdftitle={Constraining electron heat flux from reduced measurements},pdfauthor={Victor Edmonds}}
\author{Victor Edmonds\\[3pt]
  \small Final Stop Consulting LLC\\
  \small Ronin Institute for Independent Scholarship 2.0}
\date{}
\begin{document}
\maketitle
\begin{abstract}
Coronal heating estimates depend partly on electron heat transport. Different electron velocity distributions can match the same observations yet carry different heat fluxes. We ask which measurements are enough to determine heat flux without finding how electrons are distributed across all speeds and directions. The solar wind carries electrons outward from the corona and allows their velocities to be measured directly. Using one Wind measurement, we keep density, mean flow, temperature and angular summaries for each energy channel. In the fixed-bin model, electron distributions that match these quantities have heat fluxes from 3.58 to 19.85~\qunit. Keeping the correlation between electron motion along and across the magnetic field fixes the heat flux at 11.73~\qunit, matching the original bin calculation. This recovers heat flux without reconstructing the full shape of the electron velocity distribution. When the measured counts vary, retaining the required correlations leaves the same heat-flux range as direct calculation from those counts. We propose testing coronal observations in the same way to find which features of electron motion they constrain well enough to determine heat transport. Those constraints could help establish the heating requirements used to assess coronal heating mechanisms.
\end{abstract}
\section{Constraining the transport contribution to coronal heating}
\label{sec:quantity}

Assessing whether a mechanism can heat the solar corona requires knowing how much energy it must supply. That energy must balance radiation and energy carried into and out of the region by particles, including electron heat conduction \citep{withbroe_noyes1977}. Changing the assumed electron heat transport can change whether a mechanism appears to supply enough energy. We therefore need to establish what observations tell us about that transport before using it to judge a heating mechanism.

The electron velocity distribution describes how electrons are spread across speeds and directions. When observations measure only some of its properties, we face an inverse problem: work backward from those measurements to the electron distributions that could produce them. Several distributions may match the same measurements but carry different heat fluxes. Choosing one of them gives a heat-flux estimate, but we also need to know how much that estimate could change among the other distributions that fit. Determining heat flux may require much less information than determining the full electron velocity distribution.

\begin{figure}[tbp]
 \centering
 \includegraphics[width=0.92\linewidth]{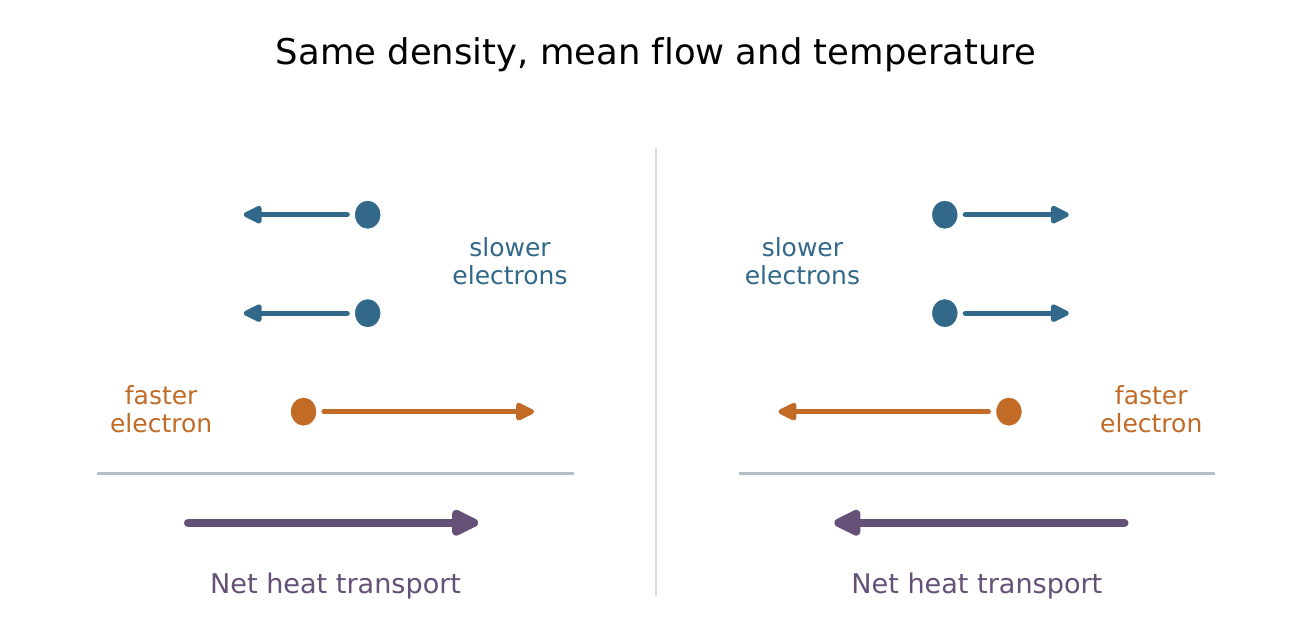}
 \caption{A schematic of opposite heat transport with equal bulk properties. On the left, two electrons move at $-v_0$ and one at $2v_0$; the right panel reverses their velocities. Both populations have the same density, zero mean flow, temperature and speed populations. Their energy fluxes point in opposite directions because the faster electron carries more energy. Arrow lengths indicate speed; purple arrows show net heat transport. This illustrates the motivation, not the Wind D4 endpoint populations.}
 \label{fig:physical}
\end{figure}

Figure~\ref{fig:physical} illustrates why equal bulk properties need not imply equal heat transport.

\subsection{The physical connection between the corona and solar wind}
\label{sec:connection}

Solar-wind electrons have traveled outward from the corona along open magnetic fields. Their starting conditions, collisions and electric and magnetic fields help shape the velocity distribution measured farther from the Sun \citep{scudder_olbert1979,cranmer_schiff2021}. Measuring these electrons gives us access to a later stage of the transport that begins in the open-field corona.

Some electron properties retain information about conditions in the corona. Using Helios and Parker Solar Probe observations, \citet{bercic2020} found that the parallel temperature of the strahl, the narrow electron population directed along the magnetic field, was approximately independent of distance from the Sun. They interpreted this temperature as evidence of the source coronal temperature and examined how scattering could change that inference. This illustrates how an electron property measured in the solar wind can inform a coronal estimate. Each such estimate needs an account of what transport preserves and what it changes.

The shape of the electron velocity distribution also matters to heat transport within the corona. Coulomb scattering depends strongly on electron energy. Electrons carrying much of the heat flux can therefore travel over different distances between deflections from the electrons that dominate density and temperature. In the stationary coronal calculation of \citet{dorelli_scudder2003}, changing the electron distributions at the boundaries leaves density and temperature profiles almost unchanged but reverses the heat flux. This result applies to their specified fields, boundaries and collision treatment. It shows how similar density and temperature profiles can be associated with different heat transport.

Wind measures electron speed and direction. We can calculate heat flux from those samples, remove some of the information, and check whether we can still recover the original value. This lets us identify what a reduced measurement must keep for a heat-flux calculation. Applying the test in the corona would require the responses of the actual coronal instruments and a model of the local transport conditions.

\subsection{An inverse test using measured electron velocities}
\label{sec:inverse}

Wind provides electron velocity measurements and calibrated moments through heat flux \citep{lin1995,wilson2026}. We create an inverse problem from one measured electron velocity distribution. We keep selected summaries of the samples and calculate the smallest and largest heat fluxes that match them. We then identify what was lost and restore it. The test asks whether this recovers heat flux without reconstructing all the details of the electron velocity distribution. We can check the result against the heat flux calculated before information was removed.

Methods for storing fewer values while preserving moments already exist. \citet{vinas_gurgiolo_2009} show how nine angular coefficients at each speed shell retain density, flow, pressure and heat flux. We compare our result with that method. Our test identifies which part of the heat-flux calculation the chosen summaries fail to determine. It also shows where assumptions about symmetry or the variation of electron speeds within a bin provide information that the summaries do not contain.

The practical question is which measured quantities a reduction should keep for its intended heat-flux calculation. Here we test whether the retained summaries support a chosen heat-flux precision and which additional quantity would help. Keeping information already present in the directional samples can then be distinguished from obtaining a new observation.

Heat flux measures random kinetic energy crossing a surface per unit area and time. Density describes the number of electrons per unit volume; temperature describes their average random energy. Heat flux also depends on direction: electrons moving in opposite directions can balance in number while carrying different energies. We calculate this third central velocity moment relative to the electron mean flow. For particle velocity $\vct v$, mean velocity $\vct U$ and random velocity $\vct c=\vct v-\vct U$, the parallel component is \citep[section~2.1]{cranmer_schiff2021}
\begin{equation}
 q_\parallel[f]=\frac{m_e}{2}\int |\vct c|^2(\vct c\cdot\vct b)
 f(\vct v)\dd^3v,\qquad
 \vct U=\frac{\int\vct v f\dd^3v}{n},\qquad n=\int f\dd^3v .
 \label{eq:target}
\end{equation}
Here $f(\vct v)$ is the electron velocity distribution, $m_e$ the electron mass, $n$ the number density and $\vct b$ the unit magnetic-field direction. Positive $q_\parallel$ points along $\vct b$. We shorten $q_\parallel$ to $q$ when the component is clear. An electron velocity distribution symmetric under $\vct c\mapsto-\vct c$ has zero heat flux. Heat flux can be nonzero when opposite-moving electrons carry different amounts of random energy.

A temperature-gradient closure estimates heat flux from the local temperature gradient. The Spitzer--Härm result gives $q_\parallel=-\kappa_\parallel\,\partial T/\partial s$, where $\kappa_\parallel$ is parallel thermal conductivity, $T$ electron temperature and $s$ distance along the field \citep{spitzer_harm1953}. It describes an electron velocity distribution kept close to local thermal equilibrium by collisions, with a small directional imbalance. Direct integration of measured electron velocities does not require this closure. We use spacecraft-frame geocentric solar ecliptic (GSE) velocities and subtract the calculated electron mean flow before evaluating heat flux.

Section~\ref{sec:reduction} calculates the heat-flux range left after retaining channel and pitch summaries. Section~\ref{sec:restoration} identifies and restores the missing correlation. Section~\ref{sec:gyrotropy} tests how symmetry and the variation of speeds within a channel affect recovery. Section~\ref{sec:source} varies the measured samples and recalculates heat flux. Section~\ref{sec:coronal} explains the next steps toward coronal constraints, followed by the conclusions in Section~\ref{sec:conclusions}. Appendix~\ref{app:notation} collects the main notation.

\section{Heat flux allowed by the retained measurements}
\label{sec:reduction}

We replace the directional samples with summaries and calculate the heat-flux range that remains. Every electron population we test must match the retained quantities. Differences in heat flux show what those quantities fail to determine.

\subsection{Measurement and reference calculation}

The example uses the EESA-L low-energy electron electrostatic analyzer of Wind's Three-Dimensional Plasma and Energetic Particle investigation (3DP; \citealp{lin1995}) for spin 58432 on 1 January 2005, with native midpoint 02:42:50.763~UTC. We call this single spin-resolved velocity distribution the measured distribution. It contains 15 energy channels and 88 look directions, or 1,320 energy--angle samples. The spacecraft-potential cut excludes the three lowest-energy channels (264 samples), leaving 12 channels with all 88 directions: 1,056 retained bins, including 46 zero-count samples.

Calibration uses spacecraft potential and geometric-factor corrections from the \nolinkurl{WIND_3DP_L3-EESALOW-MOMENTS} product \citep{wilson2026,wind_l3_data}, archived by NASA's Space Physics Data Facility (SPDF) through CDAWeb. The matched entry is zero-based CDF record 100 of the daily version-001 file. Here ``CDF record'' refers to that time-indexed entry. Appendix~\ref{app:operators} and the data supplement give the exact file, conversion routines and sample geometry.

We use the instrument's nominal energy and angular cells to define a controlled reference model. Initially, phase-space density (PSD) is constant within each retained spherical velocity bin. This fixes the unresolved structure within a cell so that we can isolate information lost when the bin populations are replaced by selected summaries. Constant PSD is a modeling assumption. Section~\ref{sec:gyrotropy} later allows speeds within a channel to vary and tests how that changes recovery.

Each unknown is the nonnegative number density integrated over a bin, called its bin mass. Analytic cell integrals give its moment contributions. The potential correction, gain factors, magnetic field and velocity support remain fixed. In this representation, the integrated density is 6.730~cm$^{-3}$, mean velocity is approximately $(-450.821,18.500,2.519)$~km\,s$^{-1}$ and scalar temperature is 13.220~eV. Temperatures are reported as energy equivalents in eV; scalar temperature is the mean of the three diagonal central-pressure components divided by density.

Our bin calculation gives 11.734~\qunit; the archived L3 EESA-L product gives 12.400~\qunit\ for the matched record \citep{wilson2026}. The published Wind procedure interpolates the samples onto a Cartesian velocity grid. Our implementation of that reconstruction and grid integration gives 12.383~\qunit\ from the same calibrated samples, 0.14\% below the archive. Direct integration of the reconstructed distribution within the same velocity domain gives 12.380~\qunit. Reconstruction and velocity-domain choices therefore account for most of the original 5.37\% difference in this controlled comparison. Density, flow and temperature still differ from the archive, and the exact production inputs and settings have not been verified (Appendix~\ref{app:comparison}). Throughout the experiment, ``recovery'' means reproducing the original bin calculation. We use its density, flow and temperature to keep the comparisons consistent. ``Nominal'' denotes values calculated from the original, unvaried samples.

\subsection{Finding the smallest and largest allowed heat fluxes}

We construct a sequence of reductions, D0--D4, to separate information about bulk properties, energy and direction. Each step tests whether more detail about one of these properties narrows the heat-flux range. These are controlled tests of which summaries to keep; Section~\ref{sec:source} also compares them with a published moment-preserving reduction.

We first retain the nominal density, mean flow and scalar temperature while reducing the directional information. Every candidate therefore uses the same mean flow in equation~\eqref{eq:target}. Density, three particle-flux components and raw scalar energy form five linear constraints, denoted D0. At fixed density and flow, retaining raw energy also retains scalar temperature.

The next steps keep more information about energy and direction. D1 adds the population in each energy channel, retaining the energy spectrum. D2 also keeps an even angular weighting in each channel that distinguishes motion along the field from motion across it. With pitch cosine $\mu=\vct v\cdot\vct b/|\vct v|$, this is the second Legendre pitch moment, weighted by $P_2(\mu)=(3\mu^2-1)/2$. D3 adds the signed weighting $\mu$, summed separately over the six lower and six higher channels, to retain a coarse energy dependence of directional imbalance. D4 keeps this first pitch moment in every channel. The six/six split is an illustrative intermediate choice; the final D4 summaries resolve every channel and do not depend on that split. All pitch angles use spacecraft-frame velocities, and the weights are averaged exactly over each angular cell. These sums still allow different electron populations around the magnetic-field direction.

Two electron populations match the retained measurements when all selected summaries agree. We find the smallest and largest heat fluxes among nonnegative populations that meet those constraints. This approach bounds the quantity of interest across models that fit the observations \citep{stark1992,evans_stark2002}. The velocity bins let us write the measurement constraints and heat flux directly in terms of electron populations.

Let $x$ contain candidate nonnegative bin masses, $m$ the nominal masses and $A_r$ all moment-weight rows retained at reduction level $r$. The subscript $r$ here labels D0--D4. At fixed flow, the heat-flux integral is the linear functional $h^Tx$, where $h$ contains the per-bin heat-flux weights and $T$ denotes transpose. The compatible set $\mathcal F_r$ and its lower and upper heat-flux bounds $q_r^-,q_r^+$ are
\begin{equation}
 \mathcal F_r=\{x\geq0:A_rx=A_rm\},\qquad
 [q_r^-,q_r^+]=
 \left[\min_{x\in\mathcal F_r}h^Tx,\ \max_{x\in\mathcal F_r}h^Tx\right].
 \label{eq:bounds}
\end{equation}
The nominal population is feasible at every level. Endpoint populations attain the bounds, with numerical checks in Appendix~\ref{app:operators}. Mixing the endpoints preserves the summaries and interpolates their heat fluxes, so every intervening value is attainable in this bin model.

\begin{figure}[htbp]
 \centering
 \includegraphics[width=\linewidth]{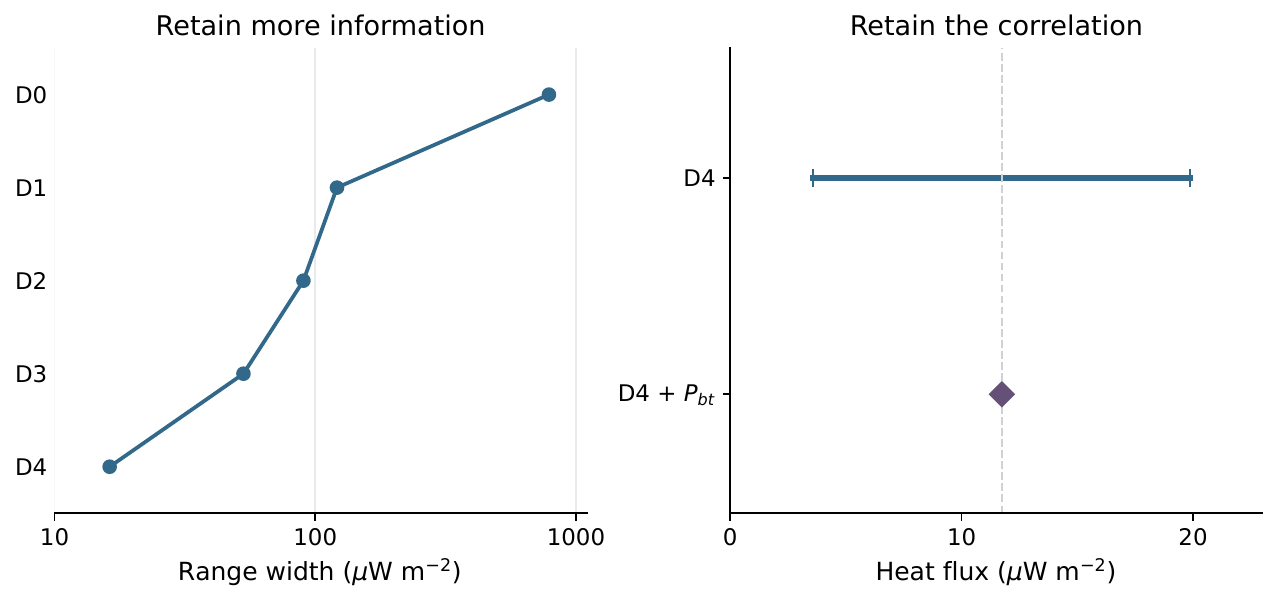}
 \caption{Information retained and heat-flux recovery for Wind/3DP EESA-L spin 58432 in the fixed-bin model. Left: heat-flux range widths under D0--D4, plotted on a logarithmic width axis. D0 keeps density, flow and scalar temperature; successive levels add channel populations, second pitch moments, coarse first pitch moments and channel-resolved first pitch moments. Right: D4 permits $[3.581,19.853]$~\qunit; adding $P_{bt}$ fixes heat flux at 11.734~\qunit. The dashed line is the original bin reference. The mean flow is fixed. These are compatible ranges, not confidence intervals.}
 \label{fig:hierarchy}
\end{figure}

Figure~\ref{fig:hierarchy} shows the contraction. D0 alone permits a width of 787.703~\qunit. Adding channel populations reduces it to 121.167~\qunit; the subsequent pitch summaries narrow it further. D4 permits $[3.581,19.853]$~\qunit. Its endpoints share density, vector flow, scalar temperature, all channel populations and the specified pitch moments. They also share parallel and perpendicular temperatures, 15.256 and 12.202~eV, because the retained rows fix those second moments.

The two endpoint populations match the same summaries but have the largest possible difference in heat flux. We next identify the term responsible for that difference and test whether retaining it fixes the heat flux.

\FloatBarrier

\section{Identifying the missing transport contribution}
\label{sec:restoration}

D4 keeps information about both speed and direction, but electron populations that match it still carry different heat fluxes. The missing term appears when we subtract the electron mean flow from spacecraft-frame velocities. The summaries use spacecraft-frame velocities; heat flux uses velocities relative to the electron flow.

Subtracting the flow across the magnetic field changes an electron's random kinetic energy even when its parallel velocity is unchanged. Heat flux multiplies that energy by parallel random velocity. The correlation between parallel and transverse motion can therefore change parallel heat flux without changing the retained pitch summaries.

Expanding the heat-flux integral shows the contributions explicitly. Define the raw second-moment tensor $\mat S_2=\int\vct v\vct v^Tf\dd^3v$ and contracted cubic moment $\vct S_3=\int|\vct v|^2\vct vf\dd^3v$. Using $\int\vct vf\dd^3v=n\vct U$ gives
\begin{equation}
\begin{split}
 q_\parallel=\frac{m_e}{2}\bigl[
  \vct b\cdot\vct S_3
  -U_b\operatorname{tr}\mat S_2
  -2\vct b^T\mat S_2\vct U
  +2n|\vct U|^2U_b\bigr],
 \qquad U_b=\vct U\cdot\vct b .
\end{split}
\label{eq:rawcentral}
\end{equation}
Here $\operatorname{tr}$ denotes the tensor trace. This expansion converts raw moments to central moments \citep{gershman_etal_2015}. \citet[section~3.3]{lavraud_larson_2016} also distinguish measured velocity coordinates from the plasma-frame pressure and heat-flux moments. Subtracting the mean flow brings the second moments into the heat-flux calculation as well as the raw cubic moment. The term $\vct b^T\mat S_2\vct U$ includes the effect of transverse flow on parallel heat flux.

\begin{figure}[tbp]
 \centering
 \includegraphics[width=0.92\linewidth]{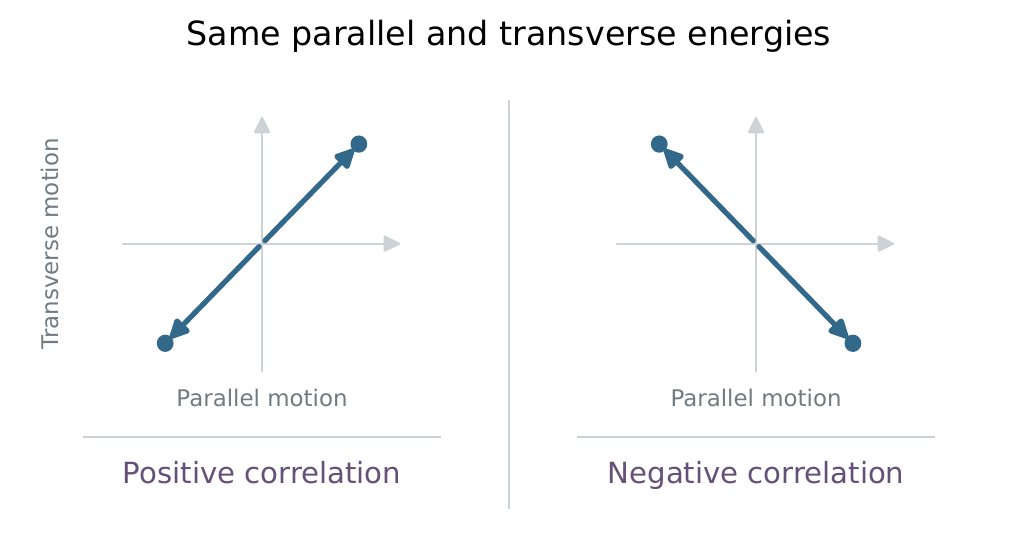}
 \caption{A schematic of the mixed pressure $P_{bt}$. Arrows show electron random velocities resolved along the magnetic field and the transverse axis $t$. Reflecting the transverse motion preserves the energies in both components and reverses their correlation. Both illustrated populations have zero mean and zero heat flux; they show why component energies do not determine mixed pressure. The Wind heat-flux recovery result is tested separately using the D4 constraints.}
 \label{fig:correlation}
\end{figure}

Write $U_b=\vct U\cdot\vct b$ and $\vct U_\perp=\vct U-U_b\vct b$. Choose the unit vector $\vct t=\vct U_\perp/|\vct U_\perp|$ along the nominal perpendicular bulk velocity, and $\vct s=\vct b\times\vct t$. Subscripts $b,t,s$ denote components along these axes; for example, $c_b=\vct c\cdot\vct b$ and $c_t=\vct c\cdot\vct t$. The central pressure tensor has components $P_{ij}=m_e\int c_ic_jf\dd^3v$, for axes $i,j\in\{b,t,s\}$. Its mixed component
\begin{equation}
 P_{bt}=m_e\int c_b c_t f\dd^3v
 \label{eq:pbt}
\end{equation}
measures the correlation between parallel and transverse random motion. Two electron populations can have the same parallel and perpendicular energies but different correlations between these velocity components. Figure~\ref{fig:correlation} illustrates this distinction.

In the original bins, D4 fixes $\vct b\cdot\vct S_3$ and $\vct b^T\mat S_2\vct b$. Constant PSD fixes the variation with speed inside each channel. The unweighted pitch summaries therefore determine these speed-weighted terms as well. D0 fixes the remaining scalar and low-order terms. For any two electron populations that match D4, subtracting equation~\eqref{eq:rawcentral} gives
\begin{equation}
 \Delta q_\parallel=-|\vct U_\perp|\,\Delta P_{bt}.
 \label{eq:missing}
\end{equation}
With pressure in eV\,cm$^{-3}$ and flow in km\,s$^{-1}$, the conversion is $\Delta q_\parallel\simeq-0.0693651\,\Delta P_{bt}$ in \qunit. The nominal perpendicular flow is about 433~km\,s$^{-1}$. At the original D4 endpoints, the pressure difference predicts the entire 16.272~\qunit\ heat-flux separation.

The channel pitch weights are independent of azimuth around $\vct b$, corresponding to spherical harmonics of order $m=0$. Here $\ell$ denotes harmonic degree and $m$ harmonic order, distinct from the bin-mass vector $m$. At fixed density and flow, changes in $P_{bt}$ come from one of the $\ell=2$, $|m|=1$ angular components that those summaries leave unmeasured \citep{vinas_gurgiolo_2009}.

Keeping this pressure component, calculated from the original directional samples, gives 11.734~\qunit\ for every electron population that matches the added constraint, reproducing the original bin calculation. We can then write the heat-flux weight as a combination of the retained measurement weights. This identity explains why all remaining populations give the same heat flux. Appendix~\ref{app:operators} gives the identity and numerical checks.

The same relation tells us how accurately the pressure component must be known. Suppose D0 and the other D4 summaries remain exact, while the added pressure component is kept within a tolerance $\pm\delta$. Two allowed pressures can then differ by at most $2\delta$. Equation~\eqref{eq:missing} limits the heat-flux range width to
\begin{equation}
 W_q\leq2|\alpha|\,\delta\simeq0.13873\,\delta,
 \label{eq:precision}
\end{equation}
where $\alpha$ is the pressure-to-flux coefficient in the units above, $\delta$ is in eV\,cm$^{-3}$ and $W_q$ is in \qunit. With the other constraints exact, we can use a desired heat-flux range width to calculate a sufficient pressure precision. The inverse calculation therefore identifies both a quantity to retain and a precision that meets the chosen heat-flux tolerance.

For example, with the other summaries exact, keeping this pressure component within $\pm7.2$~\punit\ would keep the heat-flux range width below 1~\qunit. This illustrates how to compare a proposed reduction with a chosen transport tolerance. The physical application must set that tolerance; the inverse test checks whether the retained information meets it.

Two electron populations selected using the pressure difference $P_{tt}-P_{ss}$ occupy entirely different bins but give the same recovered heat flux. We have determined heat flux without determining the full shape of the electron velocity distribution. Other calculations may still depend on the differences between those populations.

Which information is missing depends on how the samples are reduced. \citet{pulupa_etal_2014} transform Wind electron velocities before averaging over angle around the magnetic field. \citet{halekas2021} compute PSP heat flux in the solar-wind proton frame from an electron velocity distribution averaged into twelve $15^\circ$ pitch-angle sectors. In our calculation, making summaries in the electron bulk frame removes this mixed-pressure conversion term. The summaries must still preserve the speed weights needed in each channel. The angular coefficients of \citet{vinas_gurgiolo_2009} provide another way to keep the moment terms. A different reduction can lose different information and require a different correction.

\section{How symmetry and speeds within each channel affect recovery}
\label{sec:gyrotropy}

Assuming a symmetry can fix a term that the measurements leave unknown. We test gyrotropy to see which constraints come from this assumption. A gyrotropic electron velocity distribution is unchanged by rotation of random velocity around the magnetic-field direction. This rotation preserves parallel velocity and transverse speed. It does not require equal electron populations moving along and against the field, so parallel heat flux can be nonzero \citep[section~3]{cranmer_schiff2021}. Gyrotropy also makes the two transverse diagonal pressures equal and the mixed pressure components zero \citep[section~2, equations~(1)--(3)]{swisdak2016}.

\subsection{Pressure symmetry and recovery}

We first impose that pressure symmetry in the original bin representation:
\begin{equation}
 P_{bt}=P_{bs}=P_{ts}=0,\qquad P_{tt}=P_{ss}.
 \label{eq:pressuregyro}
\end{equation}
Electron populations still satisfy D0/D4 with this pressure symmetry, and both heat-flux bounds become 11.696~\qunit. Equation~\eqref{eq:missing} gives this value at $P_{bt}=0$. The original population has a small nonzero $P_{bt}$, which accounts for the difference from 11.734~\qunit. Many electron velocity distributions still give the same heat flux. Here the symmetry assumption fixes the missing pressure component and changes the recovered value.

\subsection{Allowing speeds within each channel to vary}

Constant PSD also fixes how electrons are spread across speeds within each bin. To test this assumption, we use uniformly populated rings in velocity space. These rings make the electron velocity distribution exactly gyrotropic but allow the electron speeds within each channel to vary. The original bin boundaries are centered on the spacecraft velocity origin. Rotation about the displaced bulk-frame field axis crosses these boundaries; Appendix~\ref{app:gyro} explains why exact gyrotropy requires a different representation. Each ring is uniform over
\begin{equation}
 \vct v(\phi)=\vct U+a\vct b+
 r(\vct t\cos\phi+\vct s\sin\phi),\qquad 0\leq\phi<2\pi .
 \label{eq:ring}
\end{equation}
Here $a$ is parallel random velocity, $r$ is transverse random speed and $\phi$ is gyrophase, the angle around the field axis. The radius $r$ is distinct from the earlier reduction-level index. We assign nonnegative electron populations to these rings. Any mixture of them is exactly gyrotropic, while the numbers of electrons at different speeds inside a channel can vary.

\begin{figure}[tbp]
 \centering
 \includegraphics[width=\linewidth]{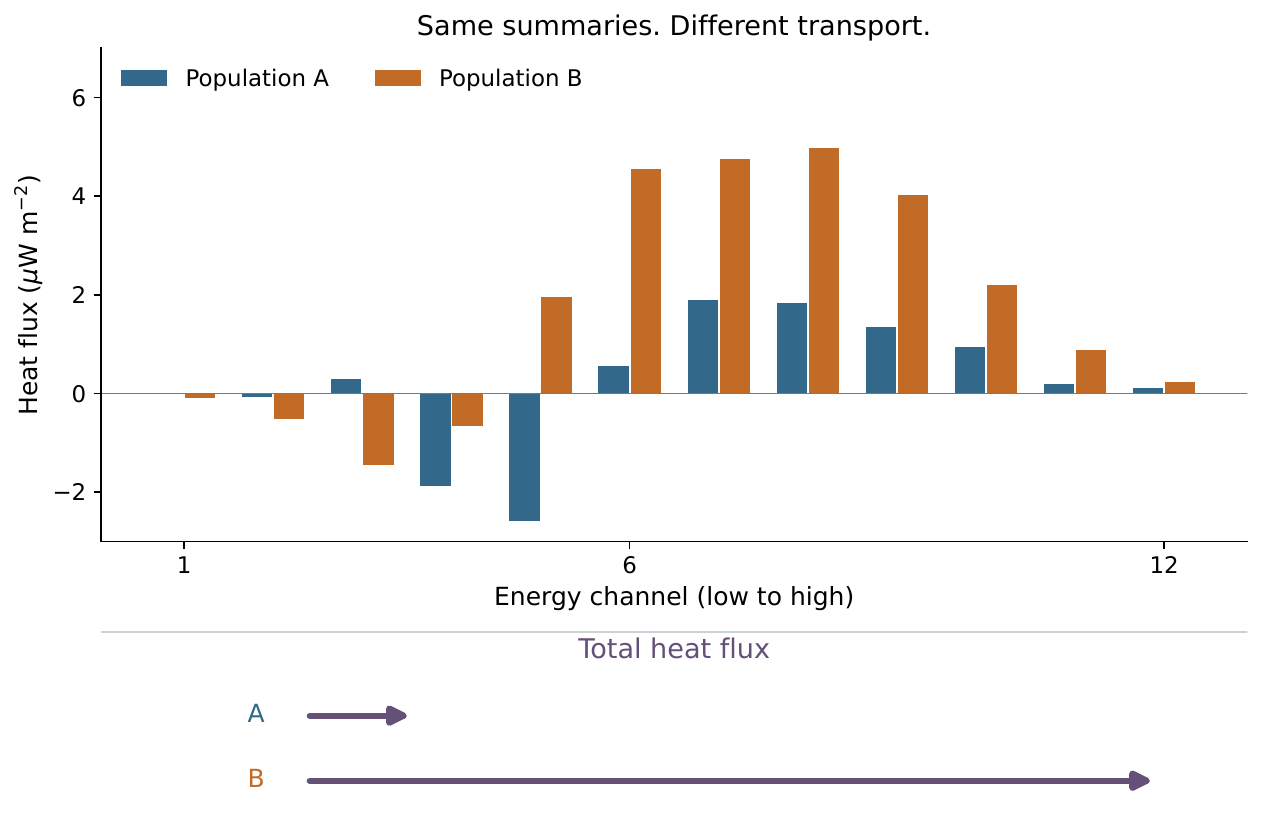}
 \caption{Heat transport from two gyrotropic ring populations matching the same D4 summaries and nominal low moments of Wind/3DP EESA-L spin 58432. Population A minimizes heat flux at 2.660~\qunit; population B maximizes it at 20.806~\qunit. The bars show signed contributions from each energy channel. Despite equal channel populations and pitch summaries, the distributions of speed and parallel motion differ. Purple arrows below show the respective totals on a common linear scale, each equal to the sum of its channel contributions.}
 \label{fig:gyro}
\end{figure}

All electrons on a ring have the same parallel random velocity and random kinetic energy. Their spacecraft-frame speeds change around the ring because spacecraft-frame velocity also includes the transverse mean flow. We calculate the fraction of each ring in each energy channel before evaluating the pitch summaries. Different mixtures of rings can match those summaries but assign different random energies to the electrons moving parallel to the field.

The ring calculation matches the original D0 and channel/pitch summaries. Small adjustments to the rounded internal channel edges make the energy channels contiguous. Appendix~\ref{app:gyro} gives the geometry and finite set of rings. We can then test whether the same summaries determine heat flux when the variation of speeds within each channel is no longer fixed.

A first pitch summary integrates $\mu f$ over a channel. The raw cubic term instead integrates $|\vct v|^3\mu f$, and the parallel second moment integrates $|\vct v|^2\mu^2f$. The extra speed weights matter. In the original bins, the fixed variation with speed links these integrals. With rings, the speeds can vary independently of gyrotropy. The method of \citet{vinas_gurgiolo_2009} keeps angular coefficients separately at each speed shell, so the required speed weights remain available when calculating the moments.

The finite set of ring mixtures gives heat fluxes from 2.660 to 20.806~\qunit. Both endpoints match the retained summaries and nominal low moments, and both have gyrotropic central pressure. Figure~\ref{fig:gyro} shows the heat flux from each channel. The channels contain the same numbers of electrons, but differences in their speeds and parallel motion produce different heat fluxes.

Keeping the two speed-weighted terms at their original bin values gives 11.696~\qunit\ for both ring bounds. Pressure symmetry fixes the mixed pressure; these two terms fix the remaining speed weights. The comparison shows why assumptions about the electron velocity distribution must be included in the inverse calculation. Those assumptions can fix heat-flux terms that the retained measurements leave unknown.

\FloatBarrier
\section{How changes in measured counts affect recovery}
\label{sec:source}

So far, the retained measurements have been exact. We now vary the measured counts to test how the recovered heat flux changes. A change in one sample can change several moments at once \citep{gershman_etal_2015}. We calculate every retained summary from the same varied samples, so these changes stay linked. We can then distinguish the heat-flux range caused by count changes from any extra range caused by discarded information.

\subsection{Varying the counts together}

We use the original bins and allow each decoded count estimate $C_i$ to change to $C_i'$, with $\max(0,C_i-\sqrt{C_i+1})\leq C_i'\leq C_i+\sqrt{C_i+1}$. We call this bounded set of count changes the count box. Calibration converts these count limits to bin-mass limits. The vector $z$ contains one allowed set of measured bin masses. Each candidate electron population $x$ must match all summaries calculated from that same $z$: $A_rx=A_rz$. Initially both $x$ and $z$ retain nominal D0. We find the smallest and largest heat fluxes over the count box. This tests sensitivity to the specified count changes; it does not define a statistical confidence region.

Keeping $P_{bt}$ reduces the heat-flux range width from 22.073 to 5.783~\qunit\ (Table~\ref{tab:angular}). Every candidate electron population then has the same heat flux as the varied samples it matches. The remaining range comes entirely from the allowed count changes. Discarding the other details of the electron velocity distribution adds no further range.

\subsection{Comparison with an existing angular reduction}

We compare the pressure correction with an existing method that keeps the moments after angular reduction. Using \citet{vinas_gurgiolo_2009} with the present cell integrals, we keep nine real angular coefficients per channel through degree $\ell=2$: 108 coefficients across 12 channels. These coefficients determine the full pressure tensor and the speed-weighted terms needed for heat flux. They give the same nominal heat flux and count-box range as D4 with $P_{bt}$. Different electron populations still match the coefficients, so the full electron velocity distribution is not determined. Appendix~\ref{app:angular} defines the angular projection and its checks.

The count-box range differs from a statistical error bar. Independent Gaussian count changes with variance $C_i+1$ give a first-order heat-flux standard deviation of 0.139~\qunit\ when D0 is held fixed (Appendix~\ref{app:covariance}). The recovered count-box half-width is 2.891~\qunit. The box allows coordinated changes across many samples, which can give a much larger range than a standard deviation under the Gaussian noise model.

\begin{table}[tbp]
 \centering
 \small
 \caption{Parallel heat flux under three reductions of Wind/3DP EESA-L spin 58432, using velocity bins with constant phase-space density (PSD). Density, vector mean flow and scalar temperature remain fixed. D4 additionally retains channel populations and first and second pitch moments in each channel; $P_{bt}$ is the mixed central pressure between parallel motion and motion along the perpendicular mean flow. The angular projection retains nine real spherical-harmonic coefficients per channel. Count-box ranges allow coordinated count changes within $C_i\pm\sqrt{C_i+1}$, clipped below at zero, with every summary taken from the same varied samples. These are bounded sensitivity ranges, not confidence intervals. All heat fluxes are in \qunit.}
 \label{tab:angular}
 \begin{tabular}{@{}lcc@{}}
 \toprule
 Retained information & Nominal heat flux & Count-box range\\
 \midrule
 D4 & $[3.581,19.853]$ & $[0.633,22.706]$\\
 D4 plus $P_{bt}$ & $11.734$ & $[8.797,14.580]$\\
 Angular projection (108 coefficients) & $11.734$ & $[8.797,14.580]$\\
 \bottomrule
 \end{tabular}
\end{table}

The weight identity explains why both methods give the same range. With the missing pressure included, each candidate electron population has the heat flux of the samples it matches: $q(x)=q(z)$. Finding the minimum and maximum directly from the allowed measured samples gives the same endpoints as the inverse calculation. Appendix~\ref{app:source} gives the proof.

Keeping $P_{bt}$ prevents the reduction from losing information already present in the directional samples. An additional independent observation could narrow the remaining range by ruling out some of the allowed samples. These are two different tasks: keep the information already measured, and obtain more information when the measurements themselves allow too wide a heat-flux range.

\begin{figure}[tbp]
 \centering
 \includegraphics[width=\linewidth]{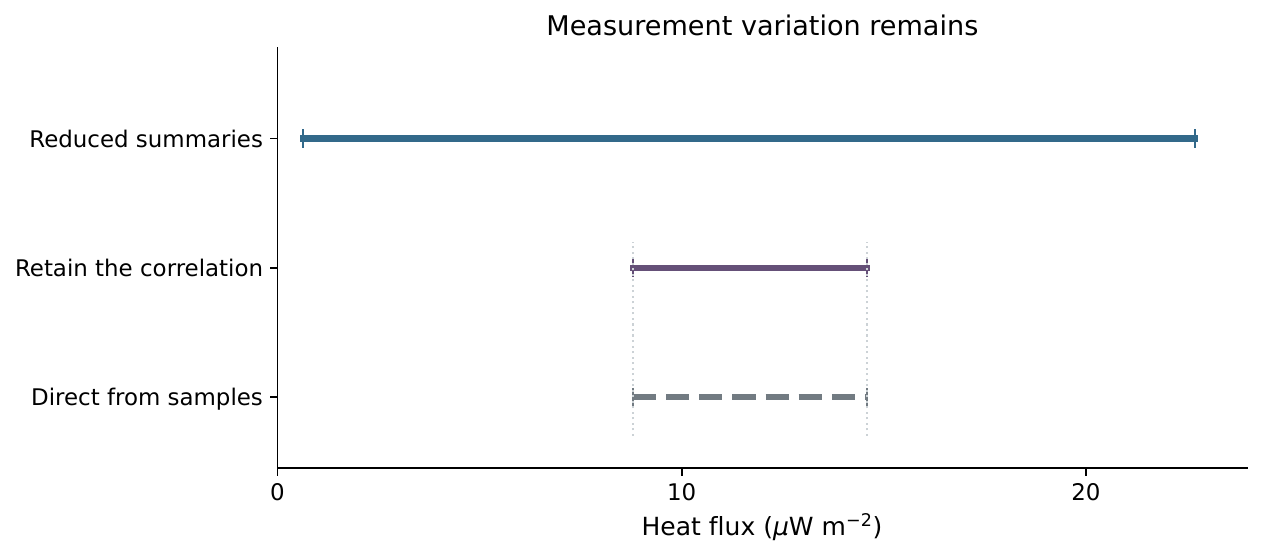}
 \caption{Separating information lost in reduction from measurement variation for Wind/3DP EESA-L spin 58432. Counts vary together within $C_i\pm\sqrt{C_i+1}$, clipped below at zero, with density, flow and scalar temperature fixed. Reduced D4 summaries permit $[0.633,22.706]$~\qunit. Keeping $P_{bt}$ narrows the range to $[8.797,14.580]$~\qunit, equal to direct integration of the allowed varied samples. The matching intervals show that the restored reduction adds no further range. These are bounded sensitivity ranges, not statistical confidence intervals.}
 \label{fig:source}
\end{figure}

\subsection{Allowing the electron mean flow to vary}

We also calculate the electron mean flow from the measured samples. It should therefore vary when the samples change. Each candidate must match the low moments of its associated samples, $A_0x=A_0z$, but those moments can now vary. Both use the same $\vct U$. A change in mean flow changes the electrons' random velocities, so central heat flux is now a nonlinear function of the samples.

Keep the original field-based axes $(\vct b,\vct t,\vct s)$, and define $U_s=\vct U\cdot\vct s$. With only $P_{bt}$ retained, the remaining heat-flux difference is
\begin{equation}
 q_\parallel(x)-q_\parallel(z)
 =-U_s(z)\,[P_{bs}(x)-P_{bs}(z)] .
 \label{eq:variableflow}
\end{equation}
The nominal flow lies in the $(\vct b,\vct t)$ plane, so $U_s=0$. Sample variation can move the flow out of that plane, bringing the second mixed pressure into the heat flux. Retaining $P_{bs}$ then recovers the heat flux for each allowed measurement again. Appendix~\ref{app:variable} gives the transformation using the fixed nominal pressure rows.

With one pressure component retained, electron populations satisfying the constraints reach heat fluxes across $[8.555,14.834]$~\qunit. With both components, they reach $[8.780,14.607]$~\qunit. The calculated outer bounds are $[8.464,14.926]$ and $[8.691,14.695]$~\qunit, respectively. Figure~\ref{fig:variableflow} in Appendix~\ref{app:variable} distinguishes the values reached by these populations from the outer bounds, which enclose every allowed heat flux. Appendix~\ref{app:variable} explains their calculation. Allowing the flow to vary reveals a second pressure component that was unnecessary when the flow direction was fixed.

\FloatBarrier
\section{Applying the test to coronal observations}
\label{sec:coronal}

We recover heat flux without reconstructing the full electron velocity distribution. In the original bins, retaining one mixed pressure removes the nominal range left by the channel and pitch summaries. Gyrotropic ring populations with free within-channel speeds again permit different heat fluxes; retaining the required speed-weighted terms fixes their heat flux. The nine-coefficient angular reduction recovers the original bin heat flux. When counts vary, retaining the required pressure components leaves only the range calculated directly from those counts.

We propose the same approach when coronal observations cannot determine the full electron velocity distribution. Start with models that predict the coronal measurements from an electron velocity distribution. Find the electron distributions that match the measurements and their uncertainties, under the physical conditions being tested. Then calculate the smallest and largest heat fluxes. If this range is too wide to assess the energy balance, identify which unmeasured electron property accounts for it and which additional observation could constrain that property. Equation~\eqref{eq:precision} gives an example: it translates a required heat-flux precision into a required pressure precision.

For measured velocity samples, the immediate use is to compare choices about what a reduction retains. One can test whether the proposed summaries meet the required heat-flux precision, then compare retaining another measured quantity with obtaining a new observation. The Wind example demonstrates the first step: retaining a pressure component removes the extra ambiguity introduced by the chosen summaries. It does not establish that an operational reduction or an existing coronal heating estimate contains that ambiguity.

The pressure correction found here depends on the spacecraft frame and the chosen channel summaries. Coronal instruments have different responses and require their own inverse calculation. The physical connection in Section~\ref{sec:connection} explains why solar-wind electrons are useful for testing the approach. Tracing a measured Wind population back to a particular coronal source would require an additional transport model. Our result covers one measured electron velocity distribution, finite sets of bins or rings, and the specified count changes. It does not tell us how often coronal observations allow different heat fluxes or how far apart those values would be.

The heating requirement also depends on how heat flux changes across the coronal region. Heat-flux divergence describes the energy deposited or removed locally; the net flux through a region's boundaries gives the corresponding total. We therefore need more than a heat-flux range at one location. Combining transport constraints across the region with the other energy terms would give a range of heating requirements. We could then ask whether that range is narrow enough to distinguish heating mechanisms or whether more observations are needed.

\section{Conclusions}
\label{sec:conclusions}

Heat flux can be determined without knowing the full shape of the electron velocity distribution. In this Wind example, keeping a measured correlation recovers the original bin heat flux, even though different electron populations still match the measurements. Changing the assumed symmetry, speeds within each channel or mean flow changes which terms must be retained. Recovering heat flux therefore depends on both the information kept and the assumptions used.

For coronal heating, the next step is to identify observations that constrain the electron properties needed for heat transport. We have demonstrated the test with measured electron speeds and directions. Repeating it with coronal instrument responses, then using the resulting heat-flux limits in the spatial energy balance, would test how tightly observations constrain the heating requirement.

\section*{Acknowledgements}
I thank NASA's Front Door program for facilitating introductions to researchers at Goddard Space Flight Center. Their perspectives helped me clarify the motivation, scope, and presentation of this work. I also thank Lynn B. Wilson III for helpful scientific discussions.
\section*{Declarations}
\noindent\textbf{Funding.} No funding was received for this work.

\noindent\textbf{Competing interests.} The author declares no competing interests.

\noindent\textbf{Data and code availability.} Numerical data and code are available through Zenodo (all versions): \url{https://doi.org/10.5281/zenodo.24115984}. The archive includes reproduction instructions and input checksums.

\noindent\textbf{AI assistance.} Codex (OpenAI) assisted with manuscript organization, analysis and computational checks, figures, and LaTeX formatting.
\begingroup
\interlinepenalty=10000
\bibliographystyle{plainnat}
\bibliography{references}
\endgroup
\clearpage
\appendix
\section{Moment operators}
\label{app:operators}

The operators below calculate the retained summaries and heat flux from the number density assigned to each velocity bin. For spherical cell $B_i$ with volume $V_i$, the original bin model is
\begin{equation}
 f_x(\vct v)=\sum_i\frac{x_i}{V_i}\mathbf1_{B_i}(\vct v),\qquad x_i\geq0 .
 \label{eq:cells}
\end{equation}
Here $\mathbf1_{B_i}$ is one inside cell $B_i$ and zero outside. Each per-mass operator is the analytic cell average of one of the weights $(1,v_a,v_av_b,|\vct v|^2v_a)$; $a,b$ denote Cartesian velocity components in this expression. The saved array includes all nine second-tensor entries. Cell masses are in cm$^{-3}$ and velocities in km\,s$^{-1}$. The calibrated inclusion mask retains 1,056 of 1,320 cells, including 46 zero-count samples. The three lowest-energy channels account for all 264 samples excluded by the potential cut. The supplied arrays store look direction before energy, with shape $[88,15]$. Velocities outside the retained support are excluded from this bin model.

The mass conversion is $M=m_e10^6/e$, where $m_e$ is the electron mass and $e$ the elementary charge. A pressure row in \punit\ uses $M$. The cubic raw-moment conversion factor $F$ gives \qunit\ through
\begin{equation}
 F=\frac12 Me10^{15}=\frac12m_e10^{21}.
\end{equation}
In these units equation~\eqref{eq:missing} is
\begin{equation}
 \Delta q_\parallel=-e10^{15}|\vct U_\perp|\,\Delta P_{bt}.
 \label{eq:unitconversion}
\end{equation}
The coefficient is $-0.06936506057145$~\qunit\ per \punit.

The nominal $P_{bt}$ is $-0.560$~\punit. The D4 endpoint pressure difference is $-234.588$~\punit, predicting the 16.272~\qunit\ heat-flux separation.

For each channel, the population, first-pitch and second-pitch rows use exact cell averages of $1,\mu,P_2(\mu)$. Within that channel the radial integration is common to its angular cells. Separability therefore relates the first-pitch rows to $\vct b\cdot\vct S_3$, and the population/second-pitch rows to $\vct b^T\mat S_2\vct b$. This relation is what changes when within-channel radial structure becomes free. D4 has 43 rows including D0 and two redundant coarse first-pitch rows, with numerical rank 38. Adding $P_{bt}$ gives 44 rows and rank 39.

If $h^T=\lambda^TA$, the retained measurements determine heat flux: its weight is a combination of the rows of $A$, with coefficients $\lambda$. With $P_{bt}$ retained, this identity has a relative maximum residual of $1.28\times10^{-15}$. The pressure results include the coefficients in the original units and their residuals. Linear programs (LPs) use HiGHS after row and objective scaling, with primal/dual solver tolerances $10^{-9}$. We accept endpoints only when the scaled equality residual and mass-bound violation are at most $2\times10^{-8}$ and the objective duality gap is at most $2\times10^{-5}$~\qunit. Checks that the intervals contain the nominal value and narrow as information is added use $10^{-5}$~\qunit\ allowances.

The primal and dual endpoint checks meet these tolerances. After the pressure component is restored, two electron populations still match the measurements and give equal heat flux, with normalized mass distance 2. For two equal-density mass vectors $x,y$, this distance is $\sum_i|x_i-y_i|/n$. A value of 2 means the populations occupy entirely different bins.

The archived comparator is zero-based record 100 of \nolinkurl{wind_3dp_l3-eesalow-moments_20050101_v001.cdf} in the L3 EESA-L moment product \citep{wilson2026,wind_l3_data}. Related Wind electron-analysis methods are described by \citet{salem2023}. The supplied calibration follows the Wind 3DP software at revision \href{https://github.com/lynnbwilsoniii/wind_3dp_pros/tree/a3100ad40afecb83f5826dbfb78b88217db453c1}{\texttt{a3100ad40a}}, including \texttt{get\_el} geometry and \texttt{convert\_esa\_units} count-to-PSD conversion. The native export already includes the $+2$~eV energy and $-5.625^\circ$ azimuth offsets. The data supplement gives the full revision identifier, source links and numerical-input checksums.

\subsection{Checking the angular reduction}
\label{app:angular}

Following the quadrature projection in \citet[section~4, equation~(9)]{vinas_gurgiolo_2009}, let $Y_\alpha$ be the nine real orthonormal harmonics of degrees $\ell=0,1,2$. For channel $j$, we store density-scaled coefficients
\begin{equation}
 \gamma_{j\alpha}=w_j a_{j\alpha}
 =\sum_{i\in j}x_i\,\frac{1}{\Omega_i}
       \int_{\Omega_i}Y_\alpha(\widehat{\vct v})\dd\Omega,
 \qquad w_j=\int_{v_j^-}^{v_j^+}v^2\dd v .
\end{equation}
Here $j$ labels an energy channel, $\alpha$ labels a harmonic, $\widehat{\vct v}=\vct v/|\vct v|$ is the unit velocity direction, $\Omega_i$ is the cell solid angle and $a_{j\alpha}$ the spectral coefficient. The radial limits $v_j^-,v_j^+$ define channel $j$, and $w_j$ is its radial volume weight. The present constant-PSD radial shape supplies the shell averages of speed $v=|\vct v|$, $v^2$ and $v^3$. Combining them with $\gamma_{j\alpha}$ recovers the raw density, first, second and contracted third moments. No distribution fit is required.

Independent 32-point angular and radial quadrature reproduces the projection and recovered raw moments. Endpoint feasibility and primal/dual checks reproduce both intervals in Table~\ref{tab:angular}. Two electron populations selected by their mass in native look index 0 have normalized mass distance 1.715 and the same recovered heat flux. They match the projected measurements but have different electron velocity distributions.

\subsection{Archived-moment comparison}
\label{app:comparison}

The bin model and the archived moment procedure assign different electron distributions to the calibrated samples. Our model holds PSD constant within each finite cell. Independent angular and radial quadrature of the central heat-flux integrand agrees with the analytic bin result to $6.85\times10^{-12}$~\qunit. This checks the integral of the specified cells.

The procedure described by \citet{wilson2026} instead uses Delaunay tetrahedra to interpolate PSD between the measured velocity points, then integrates a uniform Cartesian grid with Simpson's rule. Our Python implementation uses the public example's grid: 121 points per velocity component over $\pm15{,}000$~km\,s$^{-1}$. We fixed and saved the local velocity jitter. The original IDL code was not executed, and these settings have not been verified for the archived record. The samples use the same archived potential and geometric-factor corrections as the bin calculation.

\begin{table}[htbp]
 \centering
 \caption{Heat-flux comparison for the specified representations. Both numerical columns are in \qunit. The change is relative to the preceding row; the final row compares with the archive and is not a reproduced processing step. Each complete distribution uses its own electron mean velocity.}
 \label{tab:archive_comparison}
 \begin{tabular}{@{}lrr@{}}
 \toprule
 Representation and integration & $q_\parallel$ & Change\\
 \midrule
 Constant PSD in native cells & $11.734483096$ & ---\\
 Linear PSD over the full convex hull & $12.562161354$ & $+0.827678258$\\
 Same interpolant clipped to the cube & $12.379853423$ & $-0.182307931$\\
 Same cube, 121-point Simpson grid & $12.383317933$ & $+0.003464510$\\
 Archived L3 moment & $12.400292313$ & $+0.016974379$\\
 \bottomrule
 \end{tabular}
\end{table}

The cube is defined by $|v_x|,|v_y|,|v_z|\leq15{,}000$~km\,s$^{-1}$. Its interpolant integral uses polynomial integration after clipping the tetrahedra at this boundary, with floating-point geometry. This separates the effect of excluding the outer velocity domain from the Simpson-grid error. Both changes include the resulting change in mean velocity. Before rounding, their sum with the representation change and archive residual in Table~\ref{tab:archive_comparison} gives the original $0.665809217$~\qunit\ difference.

The local Cartesian calculation is $0.137\%$ below the archived heat flux, but its density is $3.43\%$ higher, its mean-flow vector differs by $12.62$~km\,s$^{-1}$ and its scalar temperature is $3.04\%$ lower. Thus the comparison identifies a representation difference and nearly reconciles heat flux, without reproducing the exact archived processing. It supplies no correction to the native-bin results.

Changing only the center also does not reproduce the archived heat flux. For the fixed bin population, its own mean gives 11.734~\qunit\ and the archived velocity gives 10.898~\qunit. The latter is an energy flux about the specified velocity, rather than the central heat flux of that population. Direct raw-moment expansion and the pressure-based velocity-translation relation agree. Recovery throughout the experiment refers to the original bin calculation.

\section{Exact symmetry and the gyro-ring calculation}
\label{app:gyro}

Pressure symmetry fixes selected second moments. Exact gyrotropy requires the entire electron velocity distribution to have rotational symmetry. We test these separately, then use rings to vary the electron speeds within each channel while keeping that symmetry.

\subsection{Pressure symmetry in the original bins}

The four central-pressure rows in equation~\eqref{eq:pressuregyro} are integrated analytically about nominal $\vct U$. Both D0/D4-constrained bounds equal 11.695612~\qunit. Retaining these rows still allows distinct populations with normalized mass distance 2.

\subsection{Why the original bins cannot be exactly gyrotropic}

The generator of rotation about the bulk-frame field axis is
\begin{equation}
 \vct W(\vct v)=\vct b\times(\vct v-\vct U).
\end{equation}
On an open radial cell-face patch, exact invariance requires the PSD jump to vanish whenever $\vct W$ has a nonzero normal component. For a spacecraft-centered sphere that component is proportional to $-(\vct b\times\vct U)\cdot\widehat{\vct v}$. Because $\vct U_\perp\ne0$, it is nonzero on generic open patches.

The symmetry test gives 2,112 jump constraints, one at each radial face of each retained cell, using interior angular sample points. Their matrix has rank 1,056, so the only solution sets every bin coefficient to zero. That cannot match the fixed positive density, and the LP reports no feasible solution. Thus an exactly gyrotropic electron velocity distribution cannot be represented by these spacecraft-centered step functions with nonzero density.

\subsection{Constructing rings within the energy channels}

For the uniform ring in equation~\eqref{eq:ring}, the per-particle moments are
\begin{align}
 \langle\vct v\rangle&=\vct U+a\vct b,\\
 \langle\vct v\vct v^T\rangle
 &= (\vct U+a\vct b)(\vct U+a\vct b)^T
   +\frac{r^2}{2}(\mat I-\vct b\vct b^T),\\
 q_{\parallel,\mathrm{atom}}&=F(a^2+r^2)a .
 \label{eq:ringmoments}
\end{align}
Here angle brackets denote a uniform gyrophase average and $\mat I$ is the identity tensor. The final expression is the per-particle heat-flux weight in the stated units and applies to the mixture's central target when the enforced mean velocity is $\vct U$. Each atom is a positive ring measure; no smoothness restriction is imposed on the mixture.

We specify the ring grid before solving: zero speed, 96 logarithmically spaced random speeds from 30~km\,s$^{-1}$ to the maximum saved speed plus $|\vct U|$, and 129 uniformly spaced pitch cosines. We also add rings through the 1,056 original cell-mean velocities. Removing duplicates leaves 13,441 rings before checking whether they lie within the velocity support.

Spacecraft-frame speed varies continuously around each ring. The saved channel windows have tiny gaps and overlaps because their edges were rounded independently. Requiring every point on a ring to lie inside the windows leaves 9,830 rings. Their mixtures can match D0 and D1 but cannot also match D2. Dividing overlapping channel indicators by their multiplicity leaves the gaps in place and does not restore feasibility. With those normalized indicators and D0/D1 held exact, a phase-I LP requires a maximum pitch-row change of 0.229645854 after division by channel populations.

For the contiguous-channel calculation, we replace each adjoining pair of internal edges by their arithmetic mean. The inner and outer support limits stay fixed, and the largest edge change is 0.000148612~km\,s$^{-1}$. This allows 11,419 full rings from the original grid to lie within the support. Their mixtures must still match the original D0/D4 summaries.

Channel integrals of $1,\mu,P_2(\mu)$ are evaluated by splitting each ring at channel-boundary crossings. Gauss--Legendre orders 32 and 64 agree to numerical precision. The resulting heat-flux interval is $[2.660459,20.806368]$~\qunit; both endpoints use 38 positive atoms.

Independent adaptive phase integration reproduces the retained rows and pressure symmetry, and a separate raw-to-central calculation reproduces both endpoint heat fluxes.

\subsection{Restoring the speed-weighted projections}

Append the original implied values of $B=\vct b^T\mat S_2\vct b$ and $T=\vct b\cdot\vct S_3$. The ring representation remains feasible and both bounds become 11.695612~\qunit. For the gyro-ring rows, with $U_t=|\vct U_\perp|$, the identity
\begin{equation}
\begin{split}
 q_\parallel=F\bigl[T-U_bE-2U_bB
 +(3U_b^2-U_t^2)J_b
 +(U_bU_t^2-U_b^3)n\bigr]
\end{split}
\label{eq:ringrestore}
\end{equation}
uses $E=\operatorname{tr}\mat S_2$ and $J_b=\int(\vct v\cdot\vct b)f\dd^3v$.

Centered row operations improve the conditioning of the added equations, and the fixed target row is scaled by its right-hand side. Independent checks use the original uncentered equations.

\section{Keeping all summaries tied to one varied measurement}
\label{app:source}

Each candidate electron population must match summaries calculated from one allowed set of measured samples. This keeps the changes in those summaries linked throughout the inverse calculation.

The count box in Section~\ref{sec:source} is transformed with the fixed dead-time/count-to-PSD conversion. Perturbed counts pass through float32 rounding before conversion. Energies, gain factors, potential mask and bin volumes remain fixed. Conversion validity is checked before accepting the resulting lower and upper bin-mass vectors $\ell,u$.

With fixed D0 define
\begin{align}
 Z_0&=\{z:\ell\leq z\leq u,\ A_0z=A_0m\},\\
 \mathcal G_r&=\{(x,z):x\geq0,\ z\in Z_0,\ A_0x=A_0m,\
                       A_rx=A_rz\}.
\end{align}
Here $Z_0$ is the set of allowed measured bin masses with nominal low moments, $\mathcal G_r$ is the set of compatible candidate/measurement pairs, and $A_r$ denotes the additional retained rows beyond D0 in this construction. After pressure restoration, the row-span identity gives $h^Tx=h^Tz$ on $\mathcal G_r$. Every $z\in Z_0$ also admits $(x,z)=(z,z)$, so
\begin{equation}
 \{h^Tx:(x,z)\in\mathcal G_{\mathrm D4+P_{bt}}\}
 =\{h^Tz:z\in Z_0\}.
\end{equation}
Finding the minimum and maximum heat flux directly from the allowed samples gives 8.797370904595613 and 14.579894666679420~\qunit. These are the same endpoints as the paired inverse calculation. This equality uses the original bins with D0 fixed.

\subsection{Gaussian count-noise comparison}
\label{app:covariance}

For a separate statistical comparison, assume independent Gaussian count perturbations with variance $C_i+1$. Linearize the smooth calibration before float32 rounding. If $d_i=\partial m_i/\partial C_i$, the bin-mass covariance is $\mat D=\operatorname{diag}[(C_i+1)d_i^2]$. For the dead-time conversion $f_i(C_i)=K_iC_i/(1-a_iC_i)$, the code uses $d_i=V_iK_i/(1-a_iC_i)^2$, where $a_i$ is dead time divided by dwell time and $K_i$ contains the fixed calibration factors.

Conditioning the linearized mass perturbations on $A_0\delta m=0$ gives
\begin{equation}
 \mat\Sigma_0=\mat D-\mat D A_0^T(A_0\mat D A_0^T)^{-1}A_0\mat D,
 \qquad \sigma_q^2=h^T\mat\Sigma_0h .
\end{equation}
Here $\delta m$ is a bin-mass perturbation, $\mat\Sigma_0$ its conditional covariance, $\sigma_q$ the heat-flux standard deviation, and $\operatorname{diag}$ forms a diagonal matrix. All five conditioning rows are independent. The resulting $\sigma_q$ is 0.1387969685~\qunit. A row-scaled singular-value projection and a separate covariance calculation agree. This standard deviation describes the assumed Gaussian count model; the count-box endpoints optimize over the bounded set defined above.

\section{Bounding heat flux when the mean flow varies}
\label{app:variable}

\begin{figure}[tbp]
 \centering
 \includegraphics[width=\linewidth]{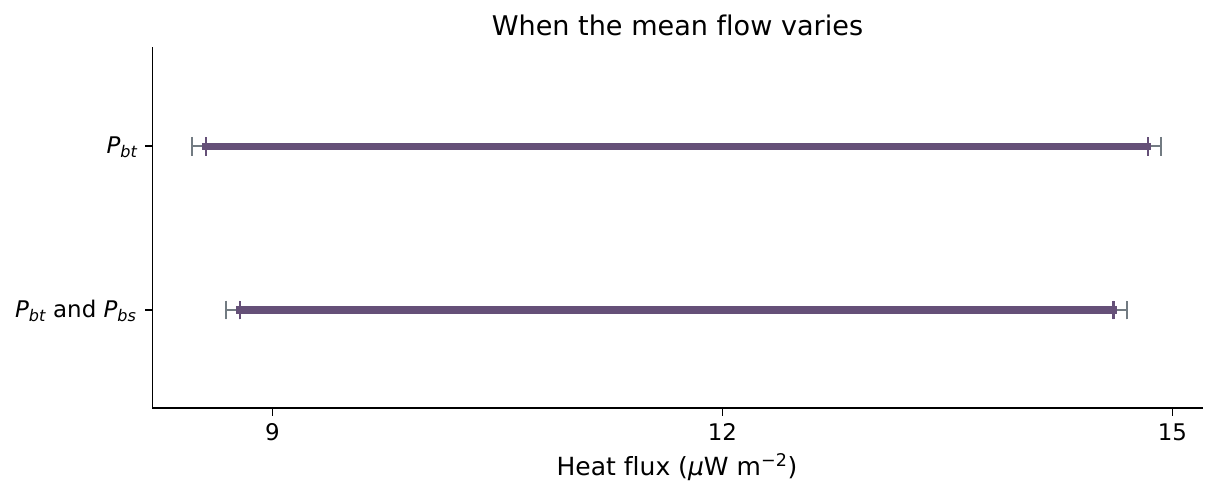}
 \caption{Heat-flux ranges when density, flow and scalar temperature vary with the allowed count samples. Candidates and their associated samples share the same low moments. Thick purple segments are feasible attained spans: $[8.555,14.834]$~\qunit\ with $P_{bt}$ retained and $[8.780,14.607]$~\qunit\ with both $P_{bt}$ and $P_{bs}$. Thin gray whiskers give conservative outer bounds, $[8.464,14.926]$ and $[8.691,14.695]$~\qunit. The horizontal scale differs from Figure~\ref{fig:source}. These are bounded sensitivity ranges, not confidence intervals.}
 \label{fig:variableflow}
\end{figure}

When $A_0x=A_0z$ varies, retain the nominal measurement axes and pressure rows. Replacing their nominal center by the common varying flow changes only shared low-moment terms, leaving candidate/source pressure differences unchanged. Equation~\eqref{eq:variableflow} follows from the raw-to-central identity.

For the source expansion, subscript 0 denotes nominal quantities. Let $\vct w=\vct v-\vct U_0$, $\vct j=\int\vct wf\dd^3v$, $\mat C=\int\vct w\vct w^Tf\dd^3v$ and $\vct d=\vct j/n$; write $d_b=\vct d\cdot\vct b$. With $\vct H(\mat C)=\vct b\operatorname{tr}\mat C+2\mat C\vct b$ and $h_0$ the nominal-centered target row, split the source heat flux into a linear part $g^Tz$ and remainder $R(z)$:
\begin{align}
 q(z)&=g^Tz+R(z),\\
 g^Tz&=h_0^Tz-F\vct H(\mat C_0)\cdot\vct j/n_0,\\
 R(z)&=-F\left[\vct H(\mat C)-\frac{n}{n_0}\vct H(\mat C_0)\right]\cdot\vct d
        +2Fn|\vct d|^2d_b .
\end{align}
Linear source quantities are bounded by sign selection over the box; linear-fractional velocity bounds use positive minimum density. Interval products give the global remainder enclosure
\begin{equation}
 R(z)\in[-0.0885447361,0.0885920475]\quad(\qunit).
\end{equation}

For the one-pressure case, LPs first bound $\Delta P_{bs}$ in $[-123.243017,121.431958]$~\punit\ and $U_s$ in $[-33.911566,33.905257]$~km\,s$^{-1}$. Sixteen equal predetermined velocity bands retain exact band inequalities and product envelopes for $U_sn=j_s$ and $U_s\Delta P_{bs}$, where $j_s=\vct j\cdot\vct s$. These envelopes follow the convex-underestimation approach of \citet{mccormick1976}. For generic scalar factors $a\in[a_l,a_h]$, $b\in[b_l,b_h]$ and product variable $p=ab$, with subscripts $l,h$ denoting lower and upper bounds, they are
\begin{align}
 p&\geq a_lb+b_la-a_lb_l,&p&\geq a_hb+b_ha-a_hb_h,\\
 p&\leq a_hb+b_la-a_hb_l,&p&\leq a_lb+b_ha-a_lb_h .
\end{align}
Central moments of feasible endpoint populations give the attained values. Combining relaxed extrema with the global remainder gives the outer enclosures. We expand the target bounds outward by $10^{-6}$~\qunit\ and apply intermediate outward guards to protect against numerical rounding.

All endpoint solves pass the primal/dual checks. Source density ranges from 6.549 to 6.911~cm$^{-3}$. The conservative temperature enclosure is $[12.746,13.705]$~eV.

\clearpage
\section{Main notation}
\label{app:notation}

The symbols below collect the main physical and measurement quantities. Definitions specific to an appendix are given with their equations. Superscript $T$ denotes transpose; $\operatorname{tr}$ denotes a tensor trace; $\Delta$ denotes a difference between compatible candidates.

\begingroup
\small
\renewcommand{\arraystretch}{1.16}
\begin{longtable}{@{}p{1.35in}>{\raggedright\arraybackslash}p{4.85in}@{}}
\toprule
Symbol & Meaning\\
\midrule
\endhead
\bottomrule
\endfoot
$f(\vct v)$ & Electron velocity distribution; PSD means phase-space density.\\
$m_e$, $e$ & Electron mass and positive elementary charge.\\
$n$ & Electron number density.\\
$\vct v$, $\vct U$, $\vct c$ & Spacecraft-frame velocity, electron mean velocity, and random velocity $\vct c=\vct v-\vct U$.\\
$\vct b$, $\vct t$, $\vct s$ & Unit axes along the magnetic field, along nominal perpendicular flow, and $\vct s=\vct b\times\vct t$.\\
$U_b$, $U_s$, $\vct U_\perp$ & Flow components along $b$ and $s$, and perpendicular flow $\vct U-U_b\vct b$.\\
$q_\parallel$, $q$ & Parallel central electron heat flux; $q$ abbreviates $q_\parallel$. Positive values point along $\vct b$.\\
$T$, $\kappa_\parallel$ & Electron temperature and parallel thermal conductivity in the gradient closure.\\
$P_{ij}$ & Central pressure component $m_e\int c_ic_jf\dd^3v$ along axes $i,j\in\{b,t,s\}$.\\
$\mu$, $P_2(\mu)$ & Spacecraft-frame pitch cosine $\vct v\cdot\vct b/|\vct v|$, and $(3\mu^2-1)/2$.\\
$\mat S_2$, $\vct S_3$ & Raw second-moment tensor $\int\vct v\vct v^Tf\dd^3v$ and contracted cubic moment $\int|\vct v|^2\vct vf\dd^3v$.\\
$x$, $m$, $z$ & Candidate, nominal and varied-measurement bin-mass vectors. Each entry is number density integrated over one velocity bin.\\
$A_r$, $A_0$, $h$ & Retained moment-weight matrix at level $r$, low-moment matrix for D0, and fixed-flow heat-flux weight vector.\\
$\mathcal F_r$, $q_r^\pm$ & Compatible candidate set and its minimum/maximum heat flux.\\
D0--D4 & Successive retained constraints: low moments, channel populations, second pitch moments, coarse first pitch moments, and channel-resolved first pitch moments.\\
$\delta$, $W_q$, $\alpha$ & Retained-pressure tolerance, heat-flux interval width, and pressure-to-flux coefficient.\\
$a$, $r$, $\phi$ & Ring parallel random velocity, transverse random speed and gyrophase. Elsewhere subscript $r$ labels the reduction level.\\
$C_i$, $C_i'$ & Nominal decoded count estimate and allowed varied count in bin $i$.\\
$\ell$, $u$ & Lower and upper bin-mass vectors for the count box.\\
$\sigma_q$ & First-order heat-flux standard deviation under the stated Gaussian count model.\\
$\ell$, $m$ (harmonics) & Spherical-harmonic degree and order, distinct from count-box vector $\ell$ and nominal mass vector $m$.\\
\end{longtable}
\endgroup

\end{document}